 \documentstyle[twocolumn,aps]{revtex}
 \input epsf
 
\begin{document}
\newcommand{\ra}{\rangle }

 \newcommand{\la}{\langle }

 \newcommand{\ket}[1]{| #1 \rangle }

 \newcommand{\bra}[1]{\langle #1 | }

 \newcommand{ \ave}[2]{ \langle #1 | #2 | #1 \rangle }

 \newcommand{\amp }[2]{\langle #1|#2 \rangle }

 \newcommand{\weakv}[3]{ \frac{\langle #1|#2| #3
 \rangle}{\langle #1 | #3 \rangle }}

 \newcommand{\beq}{\begin{equation}}

 \newcommand{\eeq}{\end{equation}}

 \newcommand{\up}[1]{^{(#1)}}

 \newcommand{\upn}{^{(N)}}

 \newcommand{\upa}{\uparrow}

 \newcommand{\dwa}{\downarrow}



 \title{ Revisiting Hardy's Paradox: Counterfactual Statements, Real Measurements,  Entanglement and Weak
 Values}

 \author{\small   Yakir Aharonov$^{(a,b,c)}$,
 Alonso Botero$^{(c,d)}$,
 Sandu Popescu$^{\bf (e,f)}$, Benni Reznik$^{a}$, Jeff
 Tollaksen$^{g}$\ \\
 $\ ^{(a)}$ {\em  \small School of Physics and Astronomy, Tel
 Aviv
                       University, Tel Aviv 69978, Israel.}\\
 $\ ^{(b)}$ {\em \small     Department of Physics,
           University of South Carolina, Columbia, SC 29208.}\\
 $\ ^{(c)}$ {\em \small     Department of Physics,
           Texas A \& M  University, College Station, TX
 7784-4242, USA.}\\
 $\ ^{(d)}$ {\em \small   Centro Internacional de F\'{\i}sica, {
 Ciudad Universitaria}, Bogot\'{a}, Colombia.}\\
 $\ ^{(e)}$ {\em  \small H.H. Wills Physics Laboratory,
 University of Bristol, Tyndall Avenue, Bristol BS8 1TL, UK}\\
 $\ ^{(f)}$ {\em \small BRIMS, Hewlett-Packard Laboratories,
 Stoke Gifford,
 Bristol BS12 6QZ, UK} \\
 $\ ^{(g)}$ {\em  \small Department of Physics, Boston
 University,
   Boston, MA 02215.}
 }

 \date{11 March 2001}
\maketitle
\begin{abstract}

Classical-realistic analysis of entangled systems have lead to
 retrodiction paradoxes, which ordinarily have been dismissed
 on
 the grounds of  counter-factuality.  Instead, we claim that
 such
 paradoxes  point to  a deeper logical structure inherent to
 quantum mechanics, which is naturally described in the
 language of
 weak values, and which is accessible experimentally via
 weak
 measurements. Using as an illustration, a
 gedanken-experiment due
 to Hardy\cite{hardy},  we show that there is in fact an exact
 numerical coincidence between a) a pair of classically
 contradictory assertions about the locations of an electron
 and a
 positron, and b) the results of  weak measurements of their
 location. The internal consistency of these results is due to
 the
 novel way by which quantum mechanics ``resolves" the
 paradox:
 first, by allowing for two distinguishable manifestations of
 how
 the electron and positron can be at the same location: either
 as
 single particles or as a pair; and secondly, by allowing these
 properties to take either sign. In particular, we discuss the
 experimental meaning of a {\em negative} number of
 electron-positron pairs.
\end{abstract}
A gedanken-experiment  due to
Hardy\cite{hardy} provides a
 beautiful illustration of  the sort of retrodiction ``paradoxes''
 arising in connection with quantum mechanical entanglement.
 To
 refute the possibility of Lorentz-invariant  elements of reality,
 he shows that in a two-particle Mach-Zehnder interferometer,
 realistic trajectories inferred from one particle's detection are
 in direct contradiction with the trajectories inferred from the
 other particle's detection.  Thus he derives a paradoxical
 inference in which an electron and a positron in some way
 manage
 to ``be'' and ``not to be'' at the same time and at the same
 location.

 A widespread tendency  to  ``resolve" the Hardy and similar
 paradoxes has been to point out that implicit in such
 paradoxes is
 an element of counter-factual reasoning,  namely, that the
 contradictions arise only because we make inferences that do
 not
 refer to results of actual experiments.  Had we actually
 performed
 the relevant measurements, we are told, then standard
 measurement
 theory predicts that the system would have been disrupted in
 such
 a way that no paradoxical implications would arise.
 \cite{desp}.

 In this Letter our claim is that one shouldn't be so quick in
 throwing away counter-factual reasoning; though indeed
 counter-factual statements have no observational meaning,
 such
 reasoning is actually a very good pointer towards interesting
 physical situations. We intend to show, {\it without invoking
 counter-factual reasoning}, that the apparently paradoxical
 reality implied counter-factually has in fact new, {\it
 experimentally accessible} consequences. These observable
 consequences become evident in terms of {\em weak
 measurements},
 which allow us to test - to some extent -  assertions that have
 been otherwise regarded as counter-factual.

 The main argument against counter-factual statements is that
 if we
 actually perform measurements to test them, we disturb the
 system significantly, and in such disturbed conditions no
 paradoxes
 arise. Our main point is that if one doesn't perform
 absolutely precise measurements but is willing to accept
 some
 finite accuracy, then one can limit the disturbance on the system. For example, according to Heisenberg's
 uncertainty
 relations, an absolutly precise measurement of position
 reduces the
 uncertainty in position to zero $\Delta x=0$ but produces an
 infinite uncertainty in momentum $\Delta p=\infty$. On the
 other
 hand, if we measure the position only up to some finite
 precision
 $\Delta x=\Delta$ we can limit the disturbance of momentum
 to a
 finite amount $\Delta p\geq \hbar/\Delta$. We use such limited
 disturbance measurements to experimentally test the
 paradoxes
 implied by the counter-factual statements. What we find is
 that
 the paradox is far from disappearing - the results of our
 measurements turn out to be most surprising and to show a
 strange, but very consistent structure.

The line of reasoning presented in our paper is very closely
related to the one suggested by Vaidman\cite{lev}.

\begin{figure} \epsfxsize=2truein
       \centerline{\epsffile{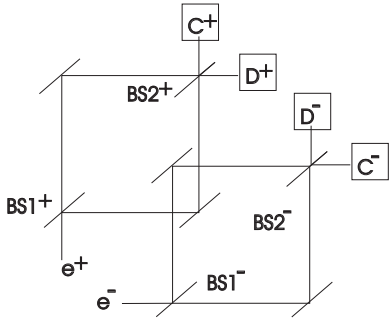}}
   \caption[]{Hardy's gedanken-experiment}
     \label{MZHardy} \end{figure}

 Let us now describe Hardy's paradox. Hardy's
 gedanken-experiment
 is a variation on the concept of interaction-free
  measurements (IFM) first suggested by Elitzur and Vaidman
 \cite{ElVaid},
  consisting of two ``superposed" Mach-Zehnder
 interferometers
  (MZI)(see Fig \ref{MZHardy}), one with a positron and one
 with an
  electron. Consider first a single interferometer, for instance
 that of
  the positron (labeled  by +). By  adjusting the arm lengths, it
  is possible to arrange specific relative phases in  the
 propagation
  amplitudes  for  paths between the beam-splitters $BS1^+$
 and $BS2^+$
  so that the positron, entering the interferometer as described
 in
  fig.1,
  can only emerge towards the
 detector $C^+$. However, the phase difference can be
 altered by the
  presence of an object, for instance in the lower arm,  in
 which case
    detector $D^+$ may be triggered. In the usual IFM setup,
 this is
  illustrated by the dramatic example of a sensitive bomb that
 absorbs
  the  particle with unit probability and subsequently explodes.
  In this way, if $D^+$ is triggered, it is then possible  to infer
   the presence of the bomb without ``touching" it, i.e., to know
 both
  that there was a bomb and that the  particle went through the
 path where
 there was no bomb.

 Now, in the double MZI setup, things are arranged so that if each
 MZI
 is considered separately,  the electron can only be detected
 at
 $C^-$ and the positron only at $C^+$. However, because
 there is
 now a region where the two particles overlap, there is also the
 possibility that they will annihilate each other. We assume
 that
 this occurs with unit probability if both particles happen to be
 in this region\footnote{Of course, we are describing here a gedanken-experiment.
 In reality the
 cross section for electron-positron annihilation is very small. We can however
 arbitrarily increase the
 annihilation probability by arranging the electron-positron to cross their paths
 many times. Also, we
 note that while we are interested in eliminating the electron-positron pair out of the
 interferometers
 when the electron and the positron happen to be in the overlapping arms, the
 actual process by
 which we do this is irrelevant for us. Annihilation is only
 one such process; scattering will do as well. For more realistic implementations see
 ``Note added" at the end of the paper.}.
   According to quantum mechanical rules, the
 presence of this interference-destroying alternative allows for
 a
 situation similar to that of the IFM in which  detectors $D^-$
 and
 $D^+$ may click in coincidence (in which case, obviously,
 there is
 no annihilation).

 But then suppose that $D^-$ and $D^+$ do click. Trying to
 ``intuitively" understand this situation leads to paradox. Based
 on the interferometers setup, we should infer from the clicking
 of
 $D^-$ that the positron must have gone through the
 overlapping
 arm; otherwise nothing would have disturbed the electron,
 and the
 electron couldn't have ended in $D^-$. Conversely, the same
 logic
 can be applied starting from the clicking of $D^+$, in which
 case
 we deduce that the electron must have also gone through the
 overlapping arm. But then they should have annihilated, and
 couldn't have reached the detectors. Hence the paradox.

 Alternatively, one could try the following line of reasoning.
 From
 the clicking of $D^-$ we infer that the positron must have
 gone
 through the overlapping arm; otherwise nothing would have
 disturbed the electron, and the electron couldn't have ended up
 in
 $D^-$. Furthermore, from the fact that there was no
 annihilation
 we also deduce that the electron must have gone through the
 non-overlapping arm. Conversely, from the clicking of $D^+$
 we
 deduce that the electron is the one which went through the
 overlapping arm and the positron went through the
 non-overlapping
 arm. But these two statements are contradictory. A paradox
 again.

 All the above statements about the positions of the electron
 and
 positron are counter-factual, i.e. we haven't actually
 measured
 the positions. Suppose however that we try to measure, say,
 the
 position of the electron, for example by inserting a detector ${\cal D}_O^-$ in
 the overlapping arm of the electron MZI. We find that, indeed, the electron is always
 in the overlapping arm - the detector ${\cal D}_O^-$ always
 clicks - in accordance with our previous counterfactual
 statements \cite{conditional}. However, ${\cal D}_O^-$
 disturbs the electron and the electron could end up in the
 $D^-$
 detector even if no positron were present! Hence, when we
 actually
 measure the position of the electron, we can no longer infer
 from
 a click at $D^-$ that a positron should have traveled through
 the
 overlapping arm of the positron MZI in order to disturb the electron. The paradox
 disappears.

 Let us now however measure the positions of the electron
 and
 positron in a more ``gentle" way, such that we do not totally
 disturb the physical observables which do not commute with
 position. To do this we will follow von Neumann's theory of
 measurement.

 Suppose we want to measure an observable $\hat A$.
 Consider a test
 particle described by the canonical position $\hat Q$ and
 conjugate
 momentum $\hat P$ which we couple the system via the
 interaction
 Hamiltonian
 \begin{equation}
 H_I=g(t)\hat P\hat A.
 \end{equation}
 The time dependent coupling constant $g(t)$ describes the
 switching ``on" and ``off" of the interaction. For an impulsive
 measurement we need the coupling to be strong and short;
 we take
 $g(t)$ to be non-zero only for a short time around the moment
 of
 interest, $t_0$ and such that
 $
 \int g(t)dt=g>0.
 $
 During the time of measurement we can neglect the effect of
 the
 free hamiltonians of the system and of the measuring device;
 the
 evolution is then governed by the interaction term and is
 given by
 the unitary operator \beq \hat{U} = e^{ -i g \hat{P} \hat{A}}.
 \eeq In the Heisenberg picture we see that the effect of the
 interaction is to shift the pointer $\hat Q$ by an amount
 proportional
 to the value of the measured observable $\hat A$, i.e. $ \hat
 Q\rightarrow
 \hat Q+g\hat A$; in effect $\hat Q$
 acts as a ``pointer" indicating the value of
 $\hat A$. The uncertainty in the reading of the pointer is given
 by
 $\Delta Q$, the initial uncertainty of $\hat Q$.

 In the Schrodinger picture the state of the measured system and
 measuring device becomes
 \beq |\Psi\ra\Psi_{MD}(Q)\rightarrow e^{ -i g \hat{P} \hat{A}
 }|\Psi\ra\Psi_{MD}(Q)=\eeq \beq=\sum_i|A=a_i\ra\la
 A=a_i|\Psi\ra\Psi_{MD}(Q-ga_i), \label{measurement}\eeq
 where
 $\Psi_{MD}(Q)$ is the initial state of the measuring device.
 For
 an ideal measurement we must know precisely the initial
 position
 of the pointer; for example $\Psi_{MD}(Q)=\delta(Q)$. Such a
 state
 is however unphysical; as a good approximation for an ideal
 measurement
 we can take a gaussian
 \beq \Psi_{MD}(Q)=\exp(-{{Q^2}\over{\Delta^2}}). \eeq When
 the
 uncertainty $\Delta$ in the initial position of the pointer is
 much smaller than the difference in the shifts of the pointer
 corresponding to different eigenvalues $a_i$, the
 measurement
 approaches an ideal measurement - the final state of the
 pointer
 (after tracing over the state of the measured system) is a
 density
 matrix representing a series of peaks, each corresponding to
 a
 different eigenvalue $a_i$, and having probability equal to
 $|\la
 A=a_i|\Psi\ra|^2$.

 However, as discussed in the introduction, we want to reduce
 the
 disturbance caused by the measurement on the measured
 system. We
 can reduce it arbitrarily by reducing the strength of the
 interaction $g$. In this regime the measurement becomes
 less
 precise since the uncertainty $\Delta$ in the initial position of
 the pointer becomes larger than the difference in the shifts of
 the pointer $ga_i$, corresponding to the different
 eigenvalues.
 Nevertheless, even in the limit of very weak interaction the
 measurement can still yield valuable information - the final
 state
 of the measuring device is almost unentangled with the
 measured
 system and approaches a gaussian centered around the {\it
 average}
 value $\bar A=\la\Psi|A|\Psi\ra$, namely
 $\Psi^{final}_{MD}\approx\exp(-{{(Q-\bar
 A)^2}\over{\Delta^2}})$.
 We need however to repeat the measurement many times to
 be able to
 locate the center.

 There is one more element we have to add.  In our example
 we are
 only interested in the results of the measuring interaction in
 the
 cases in which the electron and positron finally reached the
 detectors $D^-$ and $D^+$. To account for this we have to
 make a
 ``post-selection", i.e. to project the state (\ref{measurement})
 of the system and measuring device after the interaction on
 the
 post-selected state $|\Phi\ra$ (which in our case represents
 the
 electron and positron detected at $D^-$ and $D^+$). Thus the
 final
 state of the measuring device, given the initial state $|\Psi\ra$
 and the final state $|\Phi\ra$ is (omitting
  normalization constants)  given by
 \beq \Psi_{MD}(Q)  \to  \bra{\Phi}e^{ -i g \hat{P} \hat{A}
 }\ket{\Psi}\Psi_{MD}(Q)=\eeq
\beq=\sum_i\bra{\Phi}A=a_i\ra\la
 A=a_i|\Psi\ra\Psi_{MD}(Q-ga_i). \label{post_selected}\eeq

 As shown by Aharonov et al. \cite{a1} in the weak regime the
 effect of post-selection is very surprising. The final state of
 the measuring device (\ref{post_selected}) is \beq
 \exp({-{{(Q-g\,
 A_w)^2}\over{\Delta^2}}}), \eeq
 which describes the pointer
 shifted to a surprising value, $A_w$, called the ``weak value" of
 the
 observable $\hat A$ and given by
 \begin{equation}
 A_w = { \langle{\Phi} \vert \hat{A} \vert\Psi\rangle \over
 \langle{\Phi}\vert{\Psi}\rangle } . \label{weak} \end{equation}

 Note that in contrast to ordinary expectation values, weak
 values
 can lie outside the range of eigenvalues of $\hat{A}$ and are
 generally complex! Their real and imaginary parts are given
 by the
 corresponding effects on the  pointer $\hat Q$ and its
 conjugate $\hat P$
 respectively.

 The above behaviour of the measuring device may look
 strange
 indeed; we want to emphasize however that there is nothing
 strange
 about the measurement itself - it is an ordinary, standard
 measurement of $\hat A$,  only that the coupling $g$ with the
 measured system is made weaker. In fact, it can be shown
 that
 {\em any} external system,
 that interacts linearly
 with an observable $\hat A$ of a pre- and post-selected system,
 will react, in the limit that the coupling is sufficiently small,
 as if the value of $\hat A$ is $A_w$. (For a detailed description
 of how weak values arise, and their significance see
 \cite{tunneling}).

Finally and most importantly, in the weak regime different
measurements do not disturb each other so non-commuting variables
$\hat A$ and $\hat B$ can be measured simultaneously and they
yield the same weak values $A_w$ and $B_w$ as when measured
separately.

 Let us investigate now Hardy's paradox by using weak
 measurements.
 Let us label the arms of the interferometers as ``overlapping",
 $\rm O$, and ``non-overlapping", $\rm NO$. The state of the
 electron and positron, after passing through $BS1^-$ and
 $BS1^+$
 is

 \begin{equation}
 \frac{1}{\sqrt{2}}\left(\ \ket{\rm O}_p +
  \ket{\rm NO}_p \right)\ \times \frac{1}{\sqrt{2}}\left(\ \ket{\rm
 O}_e
 + \ket{\rm NO}_e \right)\label{initial}.
 \end{equation}
 The detectors $C^+$ and $D^+$ measure the projectors on the
 states
 $\frac{1}{\sqrt{2}}\left(\ \ket{\rm O}_p +
  \ket{\rm NO}_p \right)$ and $\frac{1}{\sqrt{2}}\left(\ \ket{\rm
 O}_p
  -
  \ket{\rm NO}_p \right)$ respectively, and similarly for the detectors $C^-$ and $D^-$. Each interferometer is
 so arranged  that
 the free propagation doesn't add any supplementary phase
 difference between the arms; if it were not for the electron-positron
 interaction, the detectors $D^-$ and $D^+$ would never click.

 We are interested in measuring the electron and positron when
 they
 traveled in the interferometers beyond the moment when they
 could
 have annihilated; we are interested in the cases when
 annihilation
 didn't occur. The state at this moment becomes

 \beq \ket{\Psi} = \frac{1}{\sqrt{3}}\ket{\rm NO}_p\ket{\rm O}_e
 +
  \frac{1}{\sqrt{3}}\ket{\rm O}_p\ket{\rm NO}_e+
  \frac{1}{\sqrt{3}}\ket{\rm NO}_p\ket{\rm
  NO}_e,\label{preselectedstate}
 \eeq which is obtained from (\ref{initial}) by projecting out the
 term $\ket{\rm O}_p\ket{\rm
 O}_e$ corresponding to annihilation.
 We take this as our initial state.

 We further restrict ourselves to the final state representing the
 simultaneous clicking of $D^-$ and $D^+$, i.e. we post-select

 \begin{equation}
 \ket{\Phi} = \frac{1}{2}\left(\ \ket{\rm NO}_p -
  \ket{\rm O}_p \right)\ \left(\ \ket{\rm
 NO}_e
 - \ket{\rm O}_e \right)\,\label{postselectedstate} .
 \end{equation}

 From the overlap of $\ket{\Psi}$ and $\ket{\Phi}$ we can
 see that this final
 possibility is indeed allowed with probability $1/12$.

What we would like to test are question such as ``Which way does
the electron go?", ``Which way does the positron go?", ``Which way
does the positron go when the electron goes through the
overlapping arm?" etc.. In other words, we would like to measure
the single-particle ``occupation" operators
 \begin{eqnarray}
 \hat N^+_{\rm NO} = \ket{\rm NO}_p\bra{\rm NO}_p \ \ & \ \
  \hat N^+_{\rm O} = \ket{\rm O}_p\bra{\rm O}_p
 \nonumber\\
 \hat N^-_{\rm NO} = \ket{\rm NO}_e\bra{\rm NO}_e \ \  &\ \
 \
 \, \hat N^-_{\rm O} = \ket{\rm O}_e\bra{\rm O}_e
 \end{eqnarray}
which tell us separately about the electron and the positron and
also the
 {\em pair} occupation operators
 \begin{eqnarray}
 \hat{N}^{+,-}_{\rm NO\, ,\rm O} = \hat N^+_{\rm NO}
 \hat N^-_{\rm O} \ \ & \ \ \hat{N}^{+,-}_{\rm O\, ,\rm NO} =
 \hat N^+_{\rm O} \hat N^-_{\rm NO} \nonumber\\
 \hat{N}^{+,-}_{\rm O\, ,\rm O} = \hat N^+_{\rm O}
 \hat N^-_{\rm O} \ \ & \ \ \hat{N}^{+,-}_{\rm NO\, ,\rm NO} =
 \hat N^+_{\rm NO} \hat N^-_{\rm NO}
 \end{eqnarray} which tell us about the simultaneous locations of
the electron and positron. We note a most important fact, which is
essential in what follows: the weak value of a product of
observables is {\it not} equal to the product of their weak
values. Hence, we have to measure the pair occupation operators
independently from the single-particle occupation
numbers\cite{lev}.

Since we will be performing weak measurements of these
observables, i.e. using probes which interact weakly with the
electron-positron system and produce only limited disturbance, we
can perform all these tests simultaneously. We will show that the
results of our measurements echo, to some extent, the
counter-factual statements, but go far beyond that. They are now
true observational statements and, if anything, they are even more
paradoxical. Indeed, using the definition of the weak value
(\ref{weak}) and the pre- and post-selected states
(\ref{preselectedstate}, \ref{postselectedstate}) we obtain

 \beq N^-_{O w}=1,~~~~~~~ N^+_{O w}=1\label{singleoverlap}\eeq

 \beq N^-_{NO w}=0,~~~~~~~ N^+_{NO w}=0\label{singlenonoverlap}\eeq

 \beq N^{+,-}_{O,O w}=0\eeq

 \beq N^{+,-}_{O,NO w}=1,~~~~~~~ N^{+,-}_{NO,O w}=1\eeq

 \beq N^{+,-}_{NO, NO w}=-1.\label{pairnonoverlap}\eeq

What do all these results tell us?

 First of all, the single-particle occupation numbers
 (\ref{singleoverlap})
 are consistent with the intuitive statements that
 ``the positron must have been in the overlapping arm
 otherwise the
 electron couldn't have ended at $D^-$" and also that ``the
 electron must have been in the overlapping arm otherwise the
 positron couldn't have ended at $D^+$". But then what
 happened to
 the fact that they could not be both in the overlapping arms
 since
 this will lead to annihilation? Quantum mechanics is
 consistent
 with this too - the pair occupation number  $N^{+,-}_{\rm O\, ,\rm O w}
 =
 0$ shows that there are zero electron-positron pairs in the
 overlapping arms!

 We also feel intuitively that ``the positron must have been in
 the
 overlapping arm otherwise the electron couldn't have ended
 at
 $D^-$, and furthermore, the electron must have gone through
 the
 non-overlapping arm since there was no annihilation". This is
 confirmed by $N^{+,-}_{\rm O\, ,\rm NO} =1$. But we also have the
 statement ``the electron must have been in the overlapping
 arm
 otherwise the positron couldn't have ended at $D^-$ and
 furthermore the positron must have gone through the
 non-overlapping arm since there was no annihilation". This is
 confirmed too, $N^{+,-}_{\rm NO\, ,\rm O w} =1$. But these two
 statements
 together are at odds with the fact that there is in fact just one
 electron-positron pair in the interferometer. Quantum
 mechanics
 solves the paradox in a remarkable way - it tells us that  $N^{+,-}_{\rm NO\, ,\rm NO w} = -1$, i.e. that there
 is
 also {\it minus} one electron-positron pair in the
 non-overlapping arms which
 brings the total down to a single pair!

Finally, the intuitive statement that ``The electron did not go
through the non-overlapping arm since it went through the
overlapping arm" is also confirmed - a weak measurement finds no
electrons in the non-overlapping arm, $N^-_{NO w}=0$. But we know
that there {\it is} one electron in the non-overlapping arm as
part of a pair in which the positron is in the overlapping arm, $
N^{+,-}_{O,NO}=1$; how is it then possible to find no electrons in
the non-overlapping arm? The answer is given by the existence of
the {\it minus} one electron-positron pair, the one with the
electron and positron in the non-overlapping arms, which
contributes a further {\it minus} one electron in the
non-overlapping arm, bringing the total number of electrons in the
non-overlapping arm to zero: \beq N^-_{NO w}=N^{+,-}_{O,NO
w}+N^{+,-}_{NO,NO w}=1-1=0.\eeq

We can now in fact go one step further. Above we have computed the
weak values by brute force. However, the weak values obey a logic
of their own which allows us to deduce them directly. We will now
follow this route since it will help us to get an intuitive
understanding of these apparently strange results. Our method is
based on two rules of behavior of weak values:

a) Suppose that between the pre-selection (preparing the initial
state) and the post-selection we perform an {\it ideal,} (von
Neumann) measurement of an observable $\hat A$, and that we
perform no other measurements between the pre- and post-selection.
Then if the outcome of this ideal measurement (given the pre- and
post-selection) is known with {\it certainty}, say $\hat A=a$ then
the weak value is equal to this particular eigenvalue, $A_w=a$.

This rule provides a direct
link to the counterfactual statements. It essentially says that all
counterfactual statements which claim that something occurs with
certainty, and which can actually be experimentally verified by
{\it separate} ideal experiments, continue to remain true when tested by weak measurements. However,
given that weak measurements do not disturb each other, all these
statements can be measured simultaneously.

b) The weak value of a sum of operators is equal to the sum of the
weak values, i.e.
\beq
\hat A=\hat B+\hat C~~~~=>~~~~~A_w=B_w+C_w
\eeq

Let us return now to Hardy's example.  As we will show, the
complete description of what occurs is encapsulated in the three
basics conterfactual statements which define the paradox:
\begin{itemize}
\item The electron is always in the overlapping arm.
\item The positron is always in the overlapping arm.
\item The electron and the positron are never both of them in the
overlapping arms.
\end{itemize}
To these counterfactual statements correspond the following {\it
observational} facts \cite{conditional}:
\begin{itemize}
\item In the cases when the electron and positron end up at
$D^-$ and $D^+$ respectively, if we measure $\hat N^-_{O}$ in an
ideal, von Neumann way, and this is the only measurement we
perform, we always find $\hat N^-_{O}=1$.
\item In the cases when the electron and positron end up at
$D^-$ and $D^+$ respectively, if we measure $\hat N^+_{O}$ in an
ideal, von Neumann way, and this is the only measurement we
perform, we always find $\hat N^+_{O}=1$.
\item In the cases when the electron and positron end up at
$D^-$ and $D^+$ respectively, if we measure $\hat N^{+,-}_{O, O}$
in an ideal, von Neumann way, and this is the only measurement we
perform, we always find $\hat N^{+,-}_{O, O}=0$.
\end{itemize}
The above statements seem paradoxical but, of course, they are
valid only if we perform the measurements separately; they do not
hold if the measurements are made simultaneously - this is the
essence of how counterfactual paradoxes are usually avoided. Rule
(a) however says that when measured weakly all these results
remain true, that is, $N^-_{O w}=1$ $N^+_{O w}=1$, $N^{+,-}_{OO
w}=0$ and can be measured simultaneously.

All other results follow from the above. Indeed, from the operator
identities

\beq \hat N^-_{O}+\hat N^-_{NO}=1\eeq

\beq \hat N^+_{O}+\hat N^+_{NO}=1\eeq

we deduce that

\beq N^-_{O w}+ N^-_{NO w}=1\eeq

\beq N^+_{O w}+ N^+_{NO w}=1\eeq which in turn imply the single
particle occupation numbers $N^-_{NO w}=0$ and $N^+_{NO w}=0$. The
operator identities

\beq \hat N^-_{O}= \hat N^{+,-}_{O,O}+\hat N^{+,-}_{NO,O}\eeq

\beq \hat N^+_{O}= \hat N^{+,-}_{O,O}+\hat N^{+,-}_{O,NO}\eeq lead
to

\beq N^-_{O w}= N^{+,-}_{O,O w}+ N^{+,-}_{NO,O w}\eeq

\beq N^+_{O w}=  N^{+,-}_{O,O w}+ N^{+,-}_{O,NO w}\eeq which in
turn imply the pair occupation numbers $N^{+,-}_{NO,O w}=1$ and
$N^{+,-}_{O,NO w}=1$.

Finally

\beq \hat N^{+,-}_{O,O}+\hat N^{+,-}_{NO,O}+\hat
N^{+,-}_{O,NO}+\hat N^{+,-}_{NO, NO}=1\eeq leads to \beq
N^{+,-}_{O,O w}+ N^{+,-}_{NO,O w}+ N^{+,-}_{O,NO w}+N^{+,-}_{O,NO
w}=1\eeq from which we obtain $N^{+,-}_{NO,NO w}=-1$.

Let us now turn to the question of how to perform the weak
measurements described above. First of all, we note that, as
discussed earlier, an important property of weak measurements, is
that what are usually mutually disturbing measurements, "commute"
in this limit, i.e., they no longer disturb each other and can be
performed simultaneously. Hence, in principle, the whole set of
predictions (\ref{singleoverlap}-\ref{pairnonoverlap}) for the
single and pair occupation numbers can be experimentally verified
simultaneously. But as we have also mentioned before, this comes
for a price - the measurements are necessarily imprecise. How
imprecise?  It can be easily seen that for the measurements
considered here (where the measured operators have only two
distinct eigenstates), the weak regime is obtained when the shift
of the pointer is smaller than the uncertainty $\Delta Q$
\cite{firstorder}. Thus in a single experiment we obtain little
information about the value of the weak values. That is, every
single measurement may yield an outcome which may be quite far
from the weak value (the spread of the outcomes around the weak
value is large). Nevertheless, by repeating the measurements (i.e.
performing a large number of independent measurements on
identically prepared systems), $A_w$ can be determined to any
desired accuracy \cite{br}. (A different, improved version of the
weak measurements will be discussed later in the paper.)

 The single particle occupation can be inferred by a weak
 measurement
 of the charge along each arm. For example by  sending
  a massive charged test particle close enough
 to the relevant path (but sufficiently distant from others)
 and then using the induced transverse momentum transfer
 as a pointer variable.
 The weakness condition is met by preparing
 the test particle to be in a localized state in the
 transverse direction,  and hence ensuring that momentum
 transfer
 is small enough.
 The measurement must be repeated many times.
 Finally, after measuring the momentum transfer in each
 experiment,
 one evaluates the mean of the result of the separate trials,
 which is taken to stand for the weak value \cite{br}.

 In each experiment one can simultaneously also measure
 the pair occupation operator by introducing
 a weak interaction between the electron and the positron.
 For instance to observe
  $\hat{N}^{+,-}_{\rm NO, \rm NO}$,
 we let the non-overlapping trajectories pass through
 two boxes, just before they arrive to the final two
 beam
 splitters. The electron and positron are temporarily captured in
the boxes and then released.
 This will not modify the experiment, provided
 that no extra phases are generated while the particles cross
 the
 boxes.
 Now suppose that the boxes are connected
 by a very rigid spring of natural length $l$.
 While the electron and positron pass
 through the boxes the relative
 deviation in the equilibrium length of the spring produced by
 the
 electrostatic force between the two boxes will be
 \beq
 \frac{\delta l}{l}  = \frac{F_{e,p}}{K l} \simeq -\frac{e^2}{K
 l^3}  N^{+,-}_{\rm NO, \rm NO} \, .
 \eeq
 where $K$ is the spring
 constant. The relative shift in the equilibrium position plays
 the
 role of the pointer variable with the
  ratio $g = e^2/K l^3 $ as a dimensionless coupling constant.
  In other words, when an electron-positron pair is present in the boxes,
  due to their electrostatic attraction the spring will be
  compressed. On the other hand, if only the electron, or only the
  positron, or none of them is present in the boxes, then here is
  no electrostatic force and the spring is left undisturbed. In
  the weak regime however, we will observe a systematic stretching
  of the spring! This is indicative of
  a negative pair occupation $N^{+,-}_{\rm NO, \rm NO}$ which
  implies an electrostatic
 {\em repulsion} between the two boxes\footnote{
 Note that since the electrostatic energy is invariant under a
 reversal of signs in
 the charges, this ``negativeness'' is not the same thing as
 charge conjugation.}.

In the above set-up the measuring devices have to be quite
imprecise in order to ensure that they do not disturb each other,
and therefore the experiment has to be repeated many times to
learn the weak values. A different version of the experiment
allows us however to measure all weak values with great precision
in one single experiment. To achieve this we send through the
interferometers a large number $\cal N$ of electron positron
pairs, one after the other. We shall now consider only the case in
which all $\cal N$ electrons end up at $D^-$ and all  $\cal N$
positrons end at $D^+$. The probability for this to happen is
exponentially small. However, when this happens, a counterfactual
reasoning similar to Hardy's original one tells us that all
electrons must have gone through the overlapping arm, all
positrons must have also gone through the overlapping arm, but
there were no electron-positron pairs in the overlapping arms.
Suppose now that we measure weakly the {\it total} number of
electrons which go through the overlapping arm. (We do this by
bringing a test charged particle near the overlapping arm, and
letting it interact with all the electrons which pass, one after
the other, through the arm.) As can easily be seen, the weak value
of the total number of electrons in the overlapping arm is
$(N^-_{O tot})_w=\cal N$. Simultaneously we use other measuring
devices to measure the weak value of the total number of positrons
and electron-positron pairs in the different arms, and so on. It
is now the case however \cite{casher}, \cite{a1}  that the
measurements no longer need to be very imprecise in order not to
significantly disturb each other. Indeed, the disturbance caused
by one measurement on the others can be reduced to an almost
negligible amount, by allowing an imprecision not greater than
$\sqrt {\cal N}$. But a $\sqrt {\cal N}$ error is negligible
compared to the total number $\cal N$ of electrons and positrons.
Thus a single experiment\footnote{We refer, of course, to a
``successful" experiment, i.e. one in which all electrons ended up
at $D^-$ and all positrons at $D^+$} is now sufficient to
determine all weak values with great precision. There is no longer
any need to average over results obtained in multiple experiments
- whenever we repeat the experiment, the measuring devices will
show the very same values, up to an insignificant spread of $\sqrt
{\cal N}$. In particular, the measuring device which measures the
total number of electron-positron pairs which went through the
non-overlapping arms shows that this number is equal to $-\cal
N\pm \sqrt {\cal N}$.

{\bf Conclusion}

In the present paper we suggested a new set of
gedanken-experiments in connection with Hardy's set-up. We find
that these experiments yield strange and surprising outcomes. As
they are experimental results, they are here to stay - they cannot
be dismissed as mere illegitimate statements about measurements
which have not been performed, as it is the case with the original
counter-factual statements. Whatever one's ultimate view about
quantum mechanics, one has to understand and explain the
significance of these outcomes.

Although the outcomes of the weak measurements suggest a story
which appears to be even stranger than Hardy's original one
(existence of a negative number of particles, etc.) the situation
is in fact far better. The weak values obey a simple, intuitive,
and, most important, {\it self-consistent} logic. This is in stark
contrast with the logic of the original counter-factual statements
which is not internally self-consistent and leads into paradoxes.
Strangeness by itself is not a problem; self-consistency is the
real issue. In this sense the logic of the weak values is similar
to the logic of special relativity: That light has the same
velocity in all reference frames is certainly highly unusual, but
everything works in a self consistent way, and because of this
special relativity is rather easy to understand. We are convinced
that, due to its self-consistency, the weak measurements logic
will lead to a deeper understanding of the nature of quantum
mechanics.

{\bf Note added} Very recently K. Moelmer has suggested a
practical way of realizing a version of the gedanken-experiment
described here, using ion trap techniques \cite{klaus}.

{\bf Acknowledgements}
 We thank A. C. Elitzur, S. Dolev and L. Vaidman for
 discussions.
 Y. A. and B.R.  acknowledge the support from grant 471/98 of
 the
 Israel Science Foundation, established by the Israel
 Academy of
 Sciences and Humanities, and NSF grant PHY-9971005.

\bigskip
\noindent {\bf Appendix}

\bigskip
In our logical derivation of the weak values we started from the
three basic statements which define Hardy's paradox, namely that
when measured separately we find with certainty that $\hat
N^-_{O}=1$, $\hat N^+_{O}=1$ and $\hat N^{+,-}_{O,O}=0$. These
three statements represent the minimal information which contains
the entire physics of the problem thus this derivation is, in a
certain sense, the most illuminating. It is useful however to give
yet another derivation.

We note that in fact we know, with certainty (in the sense of rule
(a)) quite a number of things. Apart from $\hat N^-_{O}=1$, $\hat
N^+_{O}=1$ and $\hat N^{+,-}_{O,O}=0$ we also have $\hat
N^-_{NO}=0$, $\hat N^+_{NO}=0$,  $\hat N^{+,-}_{NO,O}=1$ and $\hat
N^{+,-}_{O, NO}=1$ (see \cite{conditional}). Thus all the
corresponding weak values can be obtained directly by applying
rule (a).

Deducing the weak value of the last pair occupation number,
$N^{+,-}_{NO,NO w}$, is however more delicate. Indeed, if we
perform an ideal measurement of $\hat N^{+,-}_{NO,NO}$ we do not
obtain any certain answer. We obtain $\hat N^{+,-}_{NO,NO}=0$ with
probability $4\over5$ and $\hat N^{+,-}_{NO,NO}=1$ with
probability $1\over5$ \cite{conditional}. Rule (a) therefore does
not apply. $N^{+,-}_{NO,NO w}$ however can be deduced using the
additivity property of the weak values, together with the fact
that we know that there is only one single electron-positron pair.
Indeed, from \beq \hat N^{+,-}_{O,O}+\hat N^{+,-}_{NO,O}+\hat
N^{+,-}_{O,NO}+\hat N^{+,-}_{NO,NO}=1\eeq using additivity and the
weak values calculated above we obtain

\beq N^{+,-}_{NO,NO w}= 1-N^{+,-}_{O,O w}-N^{+,-}_{NO,O
w}-N^{+,-}_{O,NO w}=-1.\eeq



\begin{thebibliography}{99}
\bibitem{conditional}
The probability ${\cal P}(\hat A=a_i; \Psi, \Phi)$ that a von
Neumann measurement  of an observable $\hat A$ yields the value
$\hat A=a_i$ given that the initial state of the system is
$|\Psi\ra$ and given that a final measurement (performed after the
measurement of $\hat A$ ) finds the system in the state $|\Phi\ra$
is given by  ${\cal P}(\hat A=a_i; \Psi, \Phi)={{|\la\Phi| P_{\hat
A=a_i}|\Psi\ra|^2}\over{\sum_k |\la\Phi| P_{\hat
A=a_k}|\Psi\ra|^2}}$ where $P_{\hat A=a_k}$ is the projection
operator on the subspace $\hat A=a_k$, and the sum in the
denominator is taken over all eigenvalues of $\hat A$, i.e. over
all possible outcomes of the measurement of $\hat A$. Here the
numerator represents the joint probability ${\cal P}(\Phi, \hat
A=a_i; \Psi)$ that starting from $|\Psi\ra$ one obtains $\hat
A=a_i$  {\it and} that the subsequent measurement finds the system
in the state $|\Phi\ra$ while the denominator represents the
overall probability ${\cal P}(\Phi;\hat  A, \Psi)$ to find the
system in the state $|\Phi\ra$, given that $\hat A$ was measured.
See \cite{abl}.
\end{thebibliography}
 \end{document}